\documentclass[pra,superscriptaddress,twocolumn,floatfix,showpacs]{revtex4}
\usepackage{epsfig,amsbsy,amsmath,amsfonts,graphics,latexsym,subfigure}
\def\bra#1{\mathinner{\langle{#1}|}}
\def\ket#1{\mathinner{|{#1}\rangle}}

\begin{document}
\title{Decoherence-free quantum-information processing using dipole-coupled
  qubits}   
\author{Peter G. Brooke}
\email{pgb@ics.mq.edu.au}
\affiliation{Centre for Quantum Computer Technology and Department of Physics,
Macquarie University, Sydney, New South Wales, Australia}
\date{\today}
\pacs{03.67.Pp, 03.65.Yz, 03.67.Mn, 42.50.Fx}
\begin{abstract}
We propose a quantum-information processor that consists of decoherence-free
logical qubits encoded into arrays of dipole-coupled qubits. High-fidelity
single-qubit operations are performed deterministically within a
decoherence-free subsystem without leakage via global addressing of
bichromatic laser fields. Two-qubit operations are realized locally with four
physical qubits, and between separated logical qubits using linear optics. We
show how to prepare cluster states using this method. We include all
non-nearest-neighbor effects in our calculations, and we assume the qubits are
not located in the Dicke limit. Although our proposal is general to any system
of dipole-coupled qubits, throughout the paper we use nitrogen-vacancy (NV)
centers in diamond as an experimental context for our theoretical results. 
\end{abstract}
\maketitle
\section{Introduction}
In order to realize the promise of a quantum-information (QI) processor, the
inevitable decoherence-inducing effect of any system-environment interaction
must be taken into account.  One method of doing this is to encode logical
information into decoherence-free subsystems 
(DFSs)~\cite{Pal96Duan97Zan97aLidar98Knill00}.  These are regions of the
system Hilbert space that are not affected by any environment-induced
nonunitary dynamics and which under certain conditions support perfect
quantum memory.    

The general formalism of DFS theory has been applied
to a number of different physical systems, e.g., Ref.~\cite{Be00Fe02Zan98}.
In this paper, we propose a quantum-information processor that consists of
spatially separated arrays of three dipole-coupled qubits, each of which
encodes a single DF qubit. Previous work of interest includes Petrosyan and
Kurizki~\cite{Pet02}, in which the authors proposed local two-qubit operations
that used two dipole-coupled qubits.  Here, the physical qubits are also
dipole-coupled.  It is this coupling that is exploited in order to encode and
rotate a DF qubit to high fidelity, and to enable cluster-state preparation.
For the purposes of this paper, we focus on the dominant form of decoherence
at small qubit separations: {\it strong-collective} decoherence.  Fortunately,
encoding to prevent this type of decoherence is scalable to $N$-qubit
systems~\cite{Kem01}.  We show how to encode, manipulate without leakage, and 
read out QI from a DF qubit using only global control.  We then show how to
perform two-qubit operations and prepare a cluster state with the spatially
separated qubit systems.   

Complimenting theoretical progress, there have been experimental studies of
DFSs, but, although there are many ways of processing QI, deterministically
encoding, processing, and reading a single DF qubit is difficult.  In ion
traps, Kielpinski \emph{et al}.~used a DF state of two trapped ions to enable
encoded information to be stored longer than its unencoded
counterpart~\cite{Kiel01}.  This DF state has also been prepared by Kwiat
\emph{et al}.~in an optical system using parametric
down conversion~\cite{Kw00}.  More recently, the same technique 
was used to prepare a strong-collective DF qubit from four physical
qubits~\cite{Bou04}.  In liquid NMR, two physical qubits were used to
demonstrate the Deutsch-Josza algorithm~\cite{Moh03}, with the antisymmetric
collective state used to protect against decoherence.  Also in NMR, Viola
\emph{et al}.~\cite{Lo01} encoded a logical qubit into three nuclear spins in
order to protect against collective noise.  As well as realizing a similar
encoding here, we also propose a method to rotate a logical qubit to high
fidelity using only global control.  Our results are relevant to any system
that is described by a dipole-dipole interaction and, in light of recent
experimental progress, are directly applicable to nitrogen-vacancy (NV)
centers in diamond.     

NV defects in diamond have been characterized extensively~\cite{Da94}, and
recently have been used for processing quantum information.  In
Ref.~\cite{Je04}, high-fidelity one- and two-qubit operations in a single NV
defect were demonstrated. There has also been a full quantum-process tomography
of a qubit encoded in an NV defect~\cite{Ho05}.  Here, we apply results
obtained from a general master-equation analysis to the evolution of three
closely spaced dipole-coupled NV centers.  

The paper is summarized as follows.  In Sec.~\ref{sec:physys} we explicitly
define a qubit, and describe a particular unraveling of the Lindblad master
equation.  Then in Sec.~\ref{sec:prep} we propose a 
method to deterministically prepare maximally entangled states in a
three-qubit system, and give conditions that allow for preparation that is
fast relative to the decoherence rate.  In Sec.~\ref{sec:manip}, we show how to
transfer quantum information between the entangled states in a DFS,
and in Sec.~\ref{sec:rout} we use the preparation method to read out the state
of the encoded qubit.   In Sec.~\ref{sec:tqu} we propose a 
method to perform local two-qubit operations in a system of NV centers, and
then show how to prepare a cluster state with spatially separated systems of
physical qubits using linear optics.
\section{Physical system} 
\label{sec:physys}
To support the proposed encoding three qubits are required.  Although the
qubits could be realized with any system that is described by a dipole-dipole
interaction, we focus on NV centers 
in diamond.  These can be manufactured~\cite{Me05Ra05}, and consist of one
singlet ($^1$A) and two triplet states ($^3$E and $^3$A). Optical excitation
and deexcitation is  
possible only between $\text{m}_s = 0$ states~\cite{Ni03Ni03a}.  Some
deexcitation occurs to $^1\text{A}$, but the effect of this level on the
emission dynamics can be ignored~\cite{Dr99}. So, a physical qubit consists of
the electric-dipole transition $\ket{1} \equiv \ket{^3\text{E}, \text{m}_s =
  0}$, and $\ket{0} \equiv \ket{^3\text{A}, \text{m}_s = 0}$. 

For the theoretical analysis, this is equivalent to an
electric-dipole coupled two-level system with resonant frequency
$\omega_0$ and half-linewidth $\gamma$.  We assume that the spatial extent of
the vacancy is much less than the resonant wavelength $\lambda_0=2\pi/k_0$,
i.e., the qubits are effectively point dipoles.  We describe
the driving field as a classical bichromatic field that drives all three
qubits simultaneously, but due to the small qubit-qubit separations cannot
drive each qubit individually.  Regarding NV centers, we assume that since
they are closely spaced any cavity effects due to the diamond structure are
ignored, and that any incident laser field on resonance with the physical
qubits is off-resonant with all other parts of the diamond lattice.  We assume
that any inhomogeneous broadening of the electronic transition is small 
compared to $\omega_0$~\cite{San06}.  For this paper, we require aligned
dipole moments.  Fortunately, NV centers have only four orientations so, in the
absence of any field, three NV centers will be aligned with probability
$\tfrac{1}{64}$~\cite{Da94}.   

Our analysis is in accordance with standard quantum optical methods for
electric dipole-coupled qubits: a quantum master equation employing the
rotating-wave and Born-Markov approximations.  We assume that retardation
effects can be ignored, which is valid provided that the separation of qubits
$i$ and $j$ (quantified by their separation vector $\boldsymbol{r}_{ij}$) has
a value of $\xi_{ij} \equiv k_0 r _{ij} \lesssim 1$ for unit vector 
$\vec{r}_{ij} = \boldsymbol{r}_{ij} /  r_{ij} $~\cite{Mil74}. The raising
operator for qubit 
$i$ is $\hat{\sigma}_{i+}=|1\rangle_i\langle 0| = \hat{\sigma}_{i-}^\dagger$
and $\hat{\sigma}_{iz}=\tfrac{1}{2}[\hat{\sigma}_{i+}, \hat{\sigma}_{i-}]$.
The transition-matrix element of the qubit is given by $\boldsymbol{d} = \,
_i\!\bra{0}\hat{\boldsymbol{d}}_i \ket{1}_i$  with the dipole operator 
$\hat{\boldsymbol{d}}_i$ for qubit $i$.  The dipoles of the three qubits are
identically oriented, with $\vec{d} \cdot \vec{r}_{ij} = \cos \alpha$ for
$\vec{d}$ the unit vector in direction $\boldsymbol{d}$.  Since all the qubits
are identical, the matrix elements of their dipole operators are equal:
$\boldsymbol{d} = \boldsymbol{d}_i \forall i$.  

\begin{figure}[tbp]
\begin{center}
\includegraphics[width=4cm,height=3.5cm]{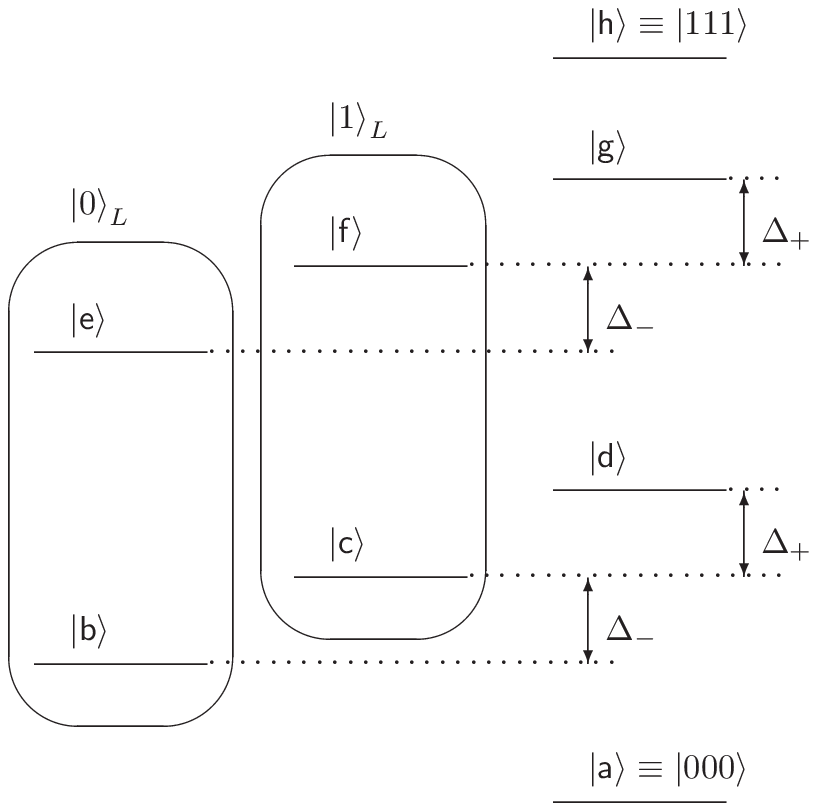}
\end{center}
\caption{\label{fig:1}Energy-level scheme of three qubits, with $
  \Delta_{\pm} = -\Delta_{13} \pm \frac{1}{2}\left(\Delta_{13} + \Omega
  \right)$, for $\Omega \equiv \sqrt{8 \Delta_{12}^2  +  \Delta_{13}^2}$.  In
  the Dicke limit~\cite{Dicke}, $\ket{\sf b} \equiv \tfrac{1}{\sqrt{6}}
  (-2\ket{001} +  \ket{010} + \ket{100})$ and $\ket{\sf c} \equiv
  \tfrac{1}{\sqrt{2}} (\ket{010} - \ket{100}) $.  The DFS is labeled $\{
  \ket{0}_L,\ket{1}_L \}$~\cite{Kem01}.}     
\end{figure}

The free evolution of three dipole-coupled qubits, including all
non-nearest-neighbor interactions, without an incident field is $(\hbar = 
1)$~\cite{Car93}     
\begin{align}
\label{eq:hnh}
\widehat{H}_{\text S} = \frac{\omega_0}{2} \sum_{i=1}^3 \hat{\sigma}_{iz} +
  \sum_{i \ne 
  j=1}^3 \Xi_{ij} \hat{\sigma}_{i+} \hat{\sigma}_{j-} - \text{i}
  \frac{\gamma}{2} \sum_{i = 1}^3  \hat{\sigma}_{i+} \hat{\sigma}_{i-}, 
\end{align}
with 
\begin{align}
\Xi_{ij} \equiv -\frac{3 \gamma}{4}
\frac{\text{e}^{\text{i}\xi_{ij}}}{\xi_{ij}^{3}} \left[ \xi_{ij}^{ 2} \sin^{2}
  \alpha  - 
  \left( 1 - \text{i}\xi_{ij} \right) \left( 1-3\cos^{2} \alpha \right) \right]
\end{align}
for $i,j = 1,2,3$, and $\xi_{ij} \equiv k_0 r _{ij}$.  By symmetry, $\Xi_{ij} =
\Xi_{ji}$, and we define $\Delta_{ij} \equiv \text{Re}\{\Xi_{ij}\}$ and $\gamma_{ij} \equiv -2 \, \text{Im} \{\Xi_{ij}\}$.  The coefficients $\Delta_{ij}$ and
$\gamma_{ij}$ correspond to the static dipole-dipole interaction and the
actual process of photon emission respectively.  We label the energy levels in
order of increasing energy: {\sf a, \ldots,h} (see Fig.~\ref{fig:1}).    

The laser has a bichromatic electric field $\boldsymbol{E}(\boldsymbol{r}) =
\boldsymbol{E}_\mu(\boldsymbol{r}) + \boldsymbol{E}_\nu(\boldsymbol{r})$, with
$\boldsymbol{E}_\mu(\boldsymbol{r})$ and $\boldsymbol{E}_\nu(\boldsymbol{r})$
the electric field amplitudes. $\boldsymbol{E}(\boldsymbol{r})$ interacts with
the three qubits via a dipole coupling and so we introduce the Rabi
frequencies $\mathcal{E}_{\mu,i} = \vec{d} \cdot \boldsymbol{E}_\mu
\text{e}^{-\text{i}   \boldsymbol{k}_\mu \cdot \boldsymbol{r}_{i}}$ and $
\mathcal{E}_{\nu,i} = 
\vec{d} \cdot \boldsymbol{E}_\nu \text{e}^{-\text{i} \boldsymbol{k}_\nu
  \cdot \boldsymbol{r}_{i}} $ for wave vectors $\boldsymbol{k}_\mu$ and
$\boldsymbol{k}_\nu$, and where qubit $i$ is situated at
$\boldsymbol{r}_i$. Within the rotating-wave approximation, the interaction
Hamiltonian is written  
\begin{align}
\widehat{H}_{\text I} =  \sum^{3}_{i=1} \mathcal{E}_{i} \widehat{\sigma}_{i-}
+ \text{H.c.},
\end{align}
for which $\mathcal{E}_{i}$ are copropagating and described
by a time-dependent bichromatic external field,
\begin{align}
 \mathcal{E}_{i}  = \mathcal{E}_{\mu,i} \text{e}^{-\text{i}\omega_\mu t } + 
\mathcal{E}_{\nu,i} \text{e}^{-\text{i}\omega_\nu t }.   
\end{align}
Due to the different positions of the qubits, the field $\mathcal{E}_{i}$
differs for distinct qubits.  The total effective Hamiltonian for the no-jump
evolution is $\widehat{H}_{\text{eff}} = \widehat{H}_{\text S} +
\widehat{H}_{\text I}$.   

The jump operators are identified by diagonalizing the relaxation matrix
$(\gamma_{ij})$~\cite{Clem03},
\begin{align}
(\gamma_{ij}) = \boldsymbol{B}^{T} \boldsymbol{\Lambda} \boldsymbol{B},
\end{align}
where $\boldsymbol{\Lambda} \equiv
\textrm{diag}\left(\lambda_1,\lambda_2,\lambda_3\right)$   
is a diagonal matrix of the eigenvalues of $(\gamma_{ij})$ and the columns of 
$\boldsymbol{B}^{T} = (b_{ij})^{T}$,
\begin{align}
\boldsymbol{b}_{l} = 
\begin{pmatrix} b_{l1} \\ b_{l2} \\ b_{l3} \end{pmatrix}
\end{align} 
are the corresponding  normalized eigenvectors.  We define
$\boldsymbol{\widehat{\Sigma}}^{\dagger} \equiv \left(
\widehat{\sigma}_{1+},\widehat{\sigma}_{2+} ,
\widehat{\sigma}_{3+} \right)$, so the jump operators are written
\begin{align}
\widehat{J}_{l} =
\sqrt{\lambda_{l}}\boldsymbol{b}^{T}_{l}\widehat{\boldsymbol{\Sigma}} 
\qquad \textrm{and} \qquad
\widehat{J}^{\dagger}_{l} = \sqrt{\lambda_{l}}
\boldsymbol{\widehat{\Sigma}}^{\dagger} \boldsymbol{b}_{l}.
\end{align}
These are quoted explicitly for three qubits in Ref.~\cite{Clem03}. In this
unraveling, the master equation is  
\begin{align}
\label{eq:me}
\dot{\hat{\rho}} = -\frac{\text{i}}{\hbar}(\widehat{H}_{\text{eff}}\hat{\rho}
- \hat{\rho} \widehat{H}^{\dagger}_{\text{eff}})
+ \sum_{i=1}^{3} 
\widehat{J}_{i}\hat{\rho}\widehat{J}_{i}^{\dagger}, 
\end{align}
which corresponds to the standard Lehmberg master equation for three
qubits~\cite{Bela69Lehm70iLehm70iiArg70} and is simply the Lindblad master 
equation~\cite{Lind}.  This unraveling is useful for analyzing preparation,
manipulation, and read out of DFS-encoded quantum information in dipole-coupled
qubits.  
\section{Preparation}
\label{sec:prep}
We describe how to prepare the maximally entangled DFS state
$\ket{\sf b}$ deterministically.  This state is the lower state of
$\ket{0}_L$, and is the longest-lived excited state in the eight-level system.
For the purposes of preparation, we require only a single laser field,
\begin{align}
\label{eq:slf}
\widehat{H}_{\text I} = \sum^{3}_{i=1} \mathcal{E}_{\mu,i}
\text{e}^{-\text{i}\omega_\mu t } \widehat{\sigma}_{i-} + \text{H.c.},
\end{align}
with frequency $\omega_\mu$, and $\boldsymbol{k}_\mu$ orthogonal to the line
joining the qubits.   
\begin{figure}[tbp]
\begin{center}
\subfigure[]{\label{fig:2a}
\includegraphics[width=3.5cm,height=2.5cm]{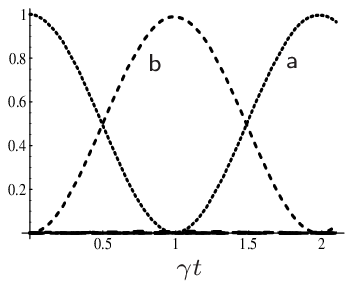}}
\subfigure[]{\label{fig:2b}
\includegraphics[width=3.5cm,height=3.0cm]{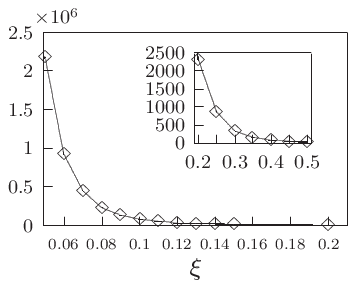}}
\end{center}
\caption{(a) \label{fig:2} Population of the levels {\sf a,\ldots,h} for
  $\xi_{12} 
  = 0.5$, $\mathcal{E}_\mu = \gamma$, $\omega_\mu =
  \tfrac{1}{2}(\Delta_{13} - \Omega )$, and $\alpha = 0$.  For these
  parameters, $F=0.988$ and the time taken for the transfer {\sf a-b}: $t_{\pi} = 0.987  
  \gamma^{-1}$.  (b) Number of population inversions possible during time
  period $\gamma_{\sf b}^{-1}$ versus qubit separation
  with $F = 0.98$ and $\alpha =0$.  For the separation proposed
  in Ref.~\cite{Me05a}, $\gamma_{\sf b} t_\pi \approx 5  \times 10^{-7}$. }      
\end{figure}
We assume that the initial state is the ground state and that the
qubits are positioned according to 
$\boldsymbol{r}_{1}= - \boldsymbol{r}$, $\boldsymbol{r}_{2}= 0$,
and $\boldsymbol{r}_{3}= \boldsymbol{r}$. Under these conditions, in the
interaction picture with respect to the Hermitian part of $\widehat{H}_{\text
  S }$, the
effective coupling between states $\ket{\sf a}$ and $\ket{\sf b}$ in the
collective basis is  
\begin{align}
\label{eq:ec}
\mathcal{E}_{\text{eff}}= \frac{1}{\Omega} \sqrt{\frac{\Omega}{\kappa}}
\mathcal{E}_\mu( \kappa \cos \boldsymbol{k}_\mu r - 2 \Delta_{12} ),
\end{align}  
where $\mathcal{E}_\mu = |\mathcal{E}_{\mu, i}|$, $ \kappa \equiv \Omega -
\Delta_{13}$, and $\Omega$ is defined in
Fig.~\ref{fig:1}.  In Eq.~\eqref{eq:ec} we have chosen $\omega_\mu =
\tfrac{1}{2}(\Delta_{13} - \Omega )$, so that $\mathcal{E}_{\text{eff}}$ is
resonant with the {\sf a-b} transition.

To ensure $\ket{\sf b}$ is prepared to high fidelity, $\Omega
\gg \gamma$.  Due to the divergence of the dipole-dipole interaction, this is
naturally satisfied at qubit separations small compared to $\lambda_0$.  For
an NV center in diamond, the separation between qubits 
for the purposes of preparation is assumed to be $r = 50\text{nm}$,
or $\xi_{12} = 0.5$.  This is within the capabilities of present
technology.  In fact, there are proposals for $r \approx 1\text{nm}$ with the
position of the qubit known to sub-nm accuracy~\cite{Me05a}.     

The effect of the laser taking into account the full eight-level 
Hamiltonian (assuming no photon emission) has been determined numerically [see
Fig.~\ref{fig:2a}].  To quantify the success of the transfer {\sf a-b}, we use
fidelity $(F)$  as a distance measure~\cite{Niel00}.  In
order to maintain high fidelity, the Rabi frequency
$\mathcal{E}_\mu$ cannot be made arbitrarily large.  This is because the
coupling $\mathcal{E}_{\sf ad} \propto \text{e}^{\text{i} \Omega t}$.
So, increasing $\mathcal{E}_\mu$ without altering $\Omega$ means the
coupling {\sf a-d} will no longer be rapidly oscillating compared to the
coupling {\sf a-b} [Eq.~\eqref{eq:ec}].   
Note also that $\mathcal{E}_{\sf ac} \propto \text{e}^{\text{i} \Omega t}$,
but since the magnitude of the coupling {\sf a-c} is much less than the
magnitude of the coupling {\sf a-d}, {\sf a-c} is weakly coupled compared to
{\sf a-d}.   
\begin{table}[bp]
\begin{tabular}{lccc}
& $F > 0.90$ & $F > 0.95$ & $F > 0.98$ \\ 
\colrule
Rabi frequency, $\mathcal{E}_\mu$ & $\pm 20\%$  & $\pm 12.5\%$ & $\pm 6 \%$ \\ 
Detuning, $\omega_\mu$ & $\pm 5\tfrac{1}{2}\%$ & $\pm 4\%$  & $\pm
1\tfrac{1}{2}\%$ \\  
Position variance, $v$  &  & 0.08 & 0.005\\ 
\end{tabular}
\caption{\label{tab:prep} Level of control for preparation of $\ket{\sf b}$
  for exact values quoted in Fig.~\ref{fig:2a} with variance $v$.  For $F >
  0.9$, the overlap of the distribution of the position of two NV 
  centers is large, causing their order to change. This renders a value for
  $v$ meaningless.}   
\end{table}

We now focus our attention on the robustness of the preparation to variations
in Rabi frequency, detuning, and separation (see Table~\ref{tab:prep}).  We
assume the Rabi frequency is equal to some value
$\mathcal{E}$, rather than $\mathcal{E}_\mu = \gamma$.  The fidelity of the
preparation after time $t_\pi$---the time taken for population
inversion---remains high, even for $\mathcal{E}/\mathcal{E}_\mu =
\tfrac{3}{4}$.  Thus, preparation is robust to small variations in 
$\mathcal{E}_\mu$.  For detuning,  we assume the laser frequency equals
some value $\omega$, rather than $\omega_\mu$.  For separation, we assume that
in practice there will always be some uncontrollable variation in qubit
separation, so the positions of the qubits are known only to a certain error.
However, once they are positioned, they do not move.  This is peculiar to NV
centers in diamond and may not apply to other physical implementations of our
proposal (e.g., atom traps).  We assume that the probability distribution of the
position of qubit $i$ is a Gaussian with mean zero and variance $v$ in units of
$\lambda_0$ centered at $\boldsymbol{r}_i$. The position of the qubit is taken
from this distribution using Monte Carlo techniques.  Table~\ref{tab:prep}
shows the limits on separation inaccuracy, averaged over $100$ different
initial positions.  If proposal~\cite{Me05a} is implemented, then typical
variations of the NV centers will be $\sim$nm, which is $\sim 0.0015
\lambda_0$, the frequency of typical lasers can be controlled to within one
part in $10^8$, and their amplitude varies by less than $0.25\%$ every $10$
seconds~\cite{newfocus}.  So, in light of Table~\ref{tab:prep} high-fidelity
preparation in NV centers, although certainly an experimental 
challenge, is possible with present technology.
\begin{figure}[tbp]
\begin{center}
\subfigure[]{\label{fig:3a}
\includegraphics[width=3.5cm,height=2.5cm]{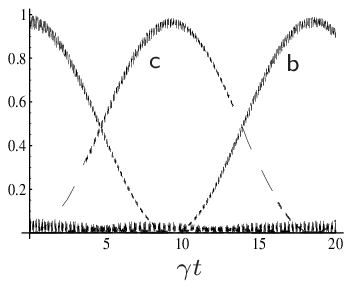}}
\subfigure[]{\label{fig:3b}
\includegraphics[width=3.5cm,height=3.0cm]{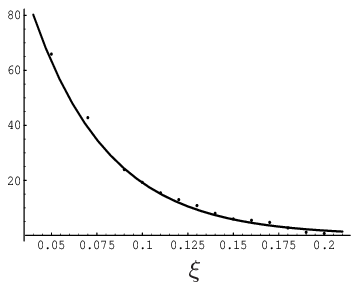}}
\end{center}
\caption{(a) \label{fig:3} Population of the levels {\sf a,\ldots,h} with
  $\xi_{12} 
  = 0.15$, $\mathcal{E}_\mu = 6 \gamma$, $\mathcal{E}_\nu = 15 \gamma$,
  $\omega_\delta = 170\gamma$, and $\alpha = \tfrac{\pi}{2}$.  For these
  parameters, $F=0.986$ and the time taken for the transfer {\sf b-c}: $t_{\pi} = 9.271
  \gamma^{-1}$.  (b) Number of qubit rotations {\sf b-c} possible during time
  period $\tfrac{1}{2}(\gamma_{\sf b} + \gamma_{\sf 
  c})^{-1}$ versus qubit separation with $F = 0.98$ 
  and $\alpha = \tfrac{\pi}{2}$. } 
\end{figure}

Before describing our method to rotate a DFS qubit, we explicitly include the
probability of decay.  The imaginary part of the eigenvalues of
Eq.~\eqref{eq:hnh} give the linewidths of the eigenstates.  The
linewidth of $\ket{\sf b}$, $\gamma_{\sf b}$, decreases with decreasing qubit
separation.  Thus, we examine the number of population inversions possible in
time period $\gamma_{\sf b}^{-1}$ with respect to physical qubit separation
[see Fig.~\ref{fig:2b}].  This improves dramatically for small separations.  
\section{Logical qubit rotation}
\label{sec:manip}
In order to rotate the encoded qubit, we use a bichromatic laser field.  With
the appropriate laser detunings this causes population 
between $\ket{\sf  b}$ and $\ket{\sf c}$ to undergo coherent oscillations,
without populating other eigenstates.  At nonzero separation the DFSs are
mixed by spontaneous emission, so in order to ignore this effect, we choose
the lowest-energy states. The laser frequencies are chosen so that 
\begin{align}
\omega_\mu = \omega_\delta \qquad \text{and} \qquad \omega_\nu =
\tfrac{1}{2}(3\Delta_{13} - \Omega ) + \omega_\delta, 
\end{align}
where $\omega_\mu$ ($\omega_\nu$) is resonant with transition {\sf b-e}({\sf
  c-e}). Similar to a two-photon Raman transition in an isolated 
three-level system,  $\omega_\mu$ and $\omega_\nu$ are detuned from resonance by
$\omega_\delta$.   The effective couplings for the transitions {\sf b-e} and {\sf c-e} are 
\begin{align}
\mathcal{E}_{\sf be} = \frac{1}{2 \Omega}\text{e}^{-\tfrac{\text{i}}{2} (\eta - 2
    \omega_\delta)t}(\text{e}^{\tfrac{\text{i}}{2} \eta t} \mathcal{E}_\mu
    + \mathcal{E}_\nu)(\kappa - 8 \Delta_{12} \cos \boldsymbol{k} r) 
\end{align}
and
\begin{align}
\mathcal{E}_{\sf ce} =&
\sqrt{\frac{2\kappa}{\Omega}}\frac{\Delta_{12}
  \eta}{\Omega^2 - \Delta_{13}(\Delta_{13} + 4
  \kappa)}  
\text{e}^{-\text{i}(\boldsymbol{k}r -
    \omega_\delta t)} \nonumber \\
&\times (1 - \text{e}^{2\text{i} \boldsymbol{k} r})
(\text{e}^{\tfrac{\text{i}}{2} \eta t} \mathcal{E}_\mu 
    + \mathcal{E}_\nu)
\end{align}
for $\boldsymbol{k}$ orthogonal to the line joining the qubits with
$\boldsymbol{k}_\mu = \boldsymbol{k}_\nu = \boldsymbol{k}$, and for $\eta
\equiv \Omega - 3 \Delta_{13}$.

Consider the three-level system {\sf b-c-e} in 
isolation, with the couplings calculated from $\widehat{H}_{\text{I}}$.  After
adiabatically eliminating $\ket{\sf e}$, the effective Rabi frequency 
between $\ket{\sf b}$ and $\ket{\sf c}$ in the collective basis and
interaction picture is   
\begin{align}
\label{eq:oef}
\mathcal{E}_{\text{eff}} =& \frac{\text{e}^{-\text{i} (\boldsymbol{k} r +
    t\eta)}\kappa^{3/2}}{4\sqrt{2}\omega_\delta (\kappa \Delta_{13} -
    2\Delta_{12}^2)^2 \Omega^{3/2}} (\text{e}^{2 \text{i} \boldsymbol{k} 
    r} - 1) \Delta_{12} \nonumber \\
&\times \eta (\text{e}^{\tfrac{\text{i}}{2} t
  \eta} \mathcal{E}_\mu + \mathcal{E}_\nu) (\mathcal{E}_\mu +
\text{e}^{\tfrac{\text{i}}{2} t \eta} \mathcal{E}_\nu)\nonumber \\
&\times (2\Delta_{12}^2 - \Delta_{13}\kappa - 2 \Delta_{12}
    \eta \cos \boldsymbol{k} r).
\end{align}
In order to calculate the detuning from resonance, $\omega_\delta$, the level
shift caused by the interaction of 
the laser with the qubits was included.  This was done numerically. We
calculate the population (assuming no 
photon emission) of all eight states for the initial state $\ket{\sf b}$ [see
Fig.~\ref{fig:3a}].  

The effective Rabi frequency cannot be made arbitrarily large: this causes
population excitation into $\ket{\sf e}$ and $\ket{\sf f}$, and, if the
increase is large enough, transitions into $\ket{\sf h}$. If
$\mathcal{E}_{\mu (\nu)} \gg \mathcal{E}_{\nu (\mu)}$, then the population
will oscillate between {\sf b-e} ({\sf c-e}).  In order for the adiabatic
elimination of the upper state {\sf e} to remain valid,
$\mathcal{E}_{\text{eff}} \ll \omega_\delta$.  We examine the robustness of
the logical qubit 
rotations to variations in Rabi frequency, detuning, and separation.  For the
Rabi frequency, we simultaneously vary $\mathcal{E}_\mu$ and $\mathcal{E}_\nu$
away from the exact values and examine $F$ at $t_\pi$, all of which are quoted
in Fig.~\ref{fig:3a}.  We examine the ratio $\mathcal{E}_\mu
\mathcal{E}_\nu/90$ for $\mathcal{E}_\mu, \mathcal{E}_\nu$ chosen close to the
values in Fig.~\ref{fig:3a} (see Table~\ref{tab:rot}).  For high-fidelity
rotations, the level of control of the Rabi frequency is high, but due to the
small separation, greater tolerance for the detuning is permitted.  Regarding
NV centers, high-fidelity rotations are possible with sub-nm position
control~\cite{Me05a}.  Note that this control assumes that once the NV
centers are located, their positions cannot be determined, and so the
frequencies of the control fields cannot be adjusted accordingly.  
\begin{table}[t]
\begin{tabular}{lccc}
& $F > 0.90$ & $F > 0.95$ & $F > 0.98$ \\ 
\colrule
Rabi frequency, $\mathcal{E}_\mu \mathcal{E}_\nu/90$ & $\pm 12\%$  & $\pm 8\%$
& $\pm 1 \%$ \\   
Detuning, $\omega_\delta$ & $\pm 28\%$ & $\pm 18\%$  & $\pm 6\%$ \\
Position variance, $v$ & $1.6 \times 10^{-3}$ & $0.8 \times 10^{-3}$ & $6
\times 10^{-6}$ \\ 
\end{tabular}
\caption{\label{tab:rot} Level of control for oscillations between {\sf b-c}
  for exact values quoted in Fig.~\ref{fig:3a} with variance $v$.}    
\end{table}

The number of rotations possible in a single spontaneous
emission lifetime enables us to see the gains obtained from using  
DFS encoding.  For the linewidth of the qubit we use $\tfrac{1}{2}(\gamma_{\sf
  b} + \gamma_{\sf c})$, which is the average linewidth of the two states.
Figure~\ref{fig:3b} shows the number of qubit rotations per spontaneous
emission lifetime.  As the separation increases, the number of rotations
decreases.  This is partly due to the increase in linewidth with increasing
separation, but mainly due to $\Delta_{\pm}$ decreasing in size.  This means
the three-level system {\sf b-c-e} can no longer be treated in 
isolation.  In fact,
\begin{align}
\lim_{\xi_{ij}\to \infty} \mathcal{E}_{\text{eff}} = 0.
\end{align}
At separations much larger than $\xi 
= 0.2$, or $20\text{nm}$, the time taken for the rotation is above that taken
for decay from {\sf b} and {\sf c}, negating the benefits of the
proposed DFS encoding for quantum-information processing.
\section{State read out}
\label{sec:rout}
This is performed in a similar way to preparation.  We exploit the
splitting of the eigenbasis in order to isolate a one-photon
transition that will fluoresce if populated.  For this purpose, a single
laser field is required [see Eq.~\eqref{eq:slf}].  Similar to
preparation, the detuning of the laser field is chosen to be 
equal to $\omega_\mu= \tfrac{1}{2}(3 \Delta_{13} + \Omega)$.  So, the (resonant)
coupling between states $\ket{\sf c}$ and $\ket{\sf g}$ is:
\begin{align}
\mathcal{E}_{\sf cg} = \frac{\text{i}}{\sqrt{2}} \sqrt{1 +
  \frac{\Delta_{13}}{\Omega}} \frac{\Delta_{12} (\Omega + 3
  \Delta_{13})\mathcal{E}_\mu \sin \boldsymbol{k} r }{2 \Delta_{12}^2 +
  \Delta_{13} (\Omega  + \Delta_{13})}.
\end{align}
If the qubit is in $\ket{1}_L$, then fluorescence will be detected, if not,
then the system will remain dark.  The decay width of state $\ket{\sf g}$ is
\begin{align}
\label{eq:glw}
\gamma_{\sf g} = \frac{1}{2}\left( 4\gamma + \gamma_{13} + \sqrt{8
  \gamma^2_{12} + \gamma^2_{13}}\right).
\end{align}    
This is superradiant with an upper bound of $4 \gamma$.  In order to read out
$\ket{0}_L$, the frequency of the laser field 
is chosen to be resonant with the transition {\sf b-g},
i.e., $\omega_\mu=\Omega$.    
\section{Two-qubit operations and cluster-state preparation}
\label{sec:tqu}
In order to perform an arbitrary sequence of quantum logic operations
two-qubit entangling operations are required~\cite{Niel00}.
Here, two methods of performing these operations are described.  The first is
a natural extension of Sec.~\ref{sec:manip}, and the second applies the
technique described in Ref.~\cite{Ch05} to a collection of systems of three closely
spaced physical qubits in order to prepare a cluster state.
\begin{figure}[t]
\begin{center}
\includegraphics[width=5.5cm,height=2.5cm]{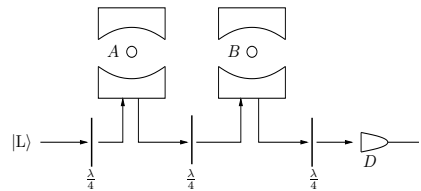}
\end{center}
\caption{\label{fig:cphase} CPHASE gate detailed
  in Ref.~\cite{Ch05} between atom $A$ and atom $B$. The $\tfrac{1}{4}$-wave 
  plates are labeled with $\tfrac{\lambda}{4}$, and the polarization detector
  with $D$.  A left-circularly polarized photon $\ket{\text L}$ is input at
  the left, and the subsequent polarization is measured at $D$.} 
\end{figure}
\subsection{Two-qubit operations}
The smallest number of physical qubits that supports two decoherence-free
qubits is four.  The decoherence-free qubits are encoded into a
decoherence-free subsystem and a decoherence-free subspace.  Using the Dicke
decomposition, the decoherence-free subspace in four qubits is~\cite{Kem01} 
\begin{align}
\ket{0}_L = &\frac{1}{2}(\ket{01}-\ket{10})(\ket{01}-\ket{10}), \\     
\ket{1}_L = &\frac{1}{\sqrt{12}}(2\ket{0011} + 2\ket{1100} -\ket{0101} 
\nonumber \\
&-\ket{1010} -\ket{0110} -\ket{1001}).
\end{align}
The Hilbert space is decomposed into irreducible
representations: $\mathcal{H}_1 \oplus \mathcal{H}_1 \oplus \mathcal{H}_3
\oplus \mathcal{H}_3 \oplus \mathcal{H}_3 \oplus \mathcal{H}_5$,  
where the subscript labels the dimension.  The
decoherence-free subspace qubit is encoded across $\mathcal{H}_1 \oplus
\mathcal{H}_1$, and the decoherence-free subsystem qubit 
across $\mathcal{H}_3 \oplus \mathcal{H}_3$.    

The controlled-phase (CPHASE) gate, $\{\ket{00}_L, \ket{01}_L, \ket{10}_L, \ket{11}_L \}$
$\to$ $\{ \ket{00}_L, \ket{01}_L, \ket{10}_L, -\ket{11}_L  \}$, is an
entangling operation.  So that two qubits will yield two logical bits of
information, the logical states are defined in pairs: $\ket{\sf b} \equiv
\ket{01}_L$, $\ket{\sf c} \equiv \ket{00}_L$, $\ket{\sf f} \equiv \ket{11}_L$,
$\ket{\sf g} \equiv \ket{10}_L$, where the Hilbert space is labeled {\sf a,b,
  \ldots,o,p} in order of increasing energy.  Using this labeling, a CPHASE
operation is performed using a $2\pi$ pulse, off resonant with transition {\sf
  f-l}, that produces a phase shift of $-1$.  Arbitrary one-qubit operations are
performed on the first qubit by resonantly coupling {\sf b-f} and {\sf c-g}
simultaneously, and on the second qubit by rotating {\sf b-c} and {\sf f-g}
using collective two-photon Raman transitions as described in
Sec.~\ref{sec:manip}.
\subsection{Cluster-state preparation} 
\label{sec:cluster}
\begin{figure}[t]
\begin{center}
\includegraphics[width=5.5cm,height=4.cm]{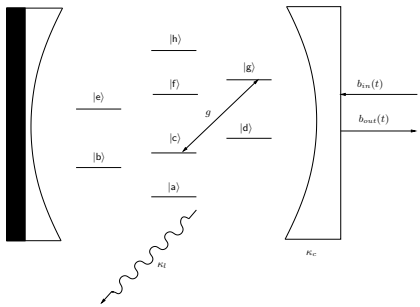}
\end{center}
\caption{\label{fig:dfcavity} The decoherence-free qubit is
  trapped in a one-sided optical cavity.  The transition {\sf c-g} is coupled
  resonantly to the right circularly polarized mode of the cavity with
  coupling constant $g$.  The cavity photon is either transmitted through the
  cavity mirror with rate $\kappa_c$ or lost with rate $\kappa_l$.
  $b_{in}(t)$ and $b_{out}(t)$ 
  denote the  input and output field operators, respectively.}     
\end{figure}
Although the previous method supports two-qubit operations, it is not
conducive to QI processing in large systems: the entangling
gate works only with isolated systems of neighboring qubits.  One method that
enables arbitrary quantum-logic operations is one-way
computation~\cite{Bri01Rau01}.  In order to generate the cluster states
required for this, we apply the method proposed in Cho and
Lee~\cite{Ch05} to a collection of spatially separated arrays of three
dipole-coupled qubits that are situated in cavities.  

In Ref.~\cite{Ch05}, the (atomic) qubit consists of the lower levels of a three-level
atom (3LA), and is situated in a one-sided optical cavity.  The right
circularly polarized mode of the cavity photon resonantly couples one of the
logical states (e.g., $\ket{1}$) to the upper state of the 3LA, but not the
other.  The phase of a right circularly polarized photon on exiting the 
cavity is unchanged if the qubit is in $\ket{1}$, otherwise the
photon, whether right or left circularly polarized, acquires a $\pi$-phase
shift.  A CPHASE gate between two separated atomic qubits can be realized
using the setup shown in Fig.~\ref{fig:cphase}.  

This system can be realized here by placing the logical qubit inside a
cavity (see Fig.~\ref{fig:dfcavity}).  The
encoded qubit is placed in a cavity on resonance with $\frac{1}{2}\left(3
\Delta_{13} + \Omega \right)$ so only the right circularly
polarized cavity photon interacts with the qubit in state $\ket{\sf c}$.  If
the qubit is in $\ket{\sf b}$, the cavity photon acquires a $\pi$-phase shift,
otherwise it does not.  Then, two-qubit operations are performed in the
same manner as described in Ref.~\cite{Ch05}.    

However, our setup has a number of further requirements.  First, the
atom-cavity coupling rate has to be fast compared to the decay time scale of
the DF state $\ket{\sf c}$. Second, $\ket{\sf g}$ is superradiant
[Eq.~\eqref{eq:glw}], so the atom-cavity coupling rate must be fast
compared to $\gamma_{\sf g}$. Fortunately, the upper bound of $\gamma_{\sf g}$
is $4\gamma$, which is small compared to $\Delta_\pm$.  Third, as well as the
probability of logical qubit decay, if there is a decay from {\sf g}, the most
probable decay channel is {\sf g-d} not {\sf g-c}, so any decay implies
information loss.  

Cluster states can be generated as follows~\cite{Ch05}. The
$1$D cluster state of $N$ qubits can be written as 
\begin{align}
\label{eq:clus}
\ket{\Psi_{N}}& = \frac{1}{\sqrt{2}}\ket{\phi_0}_{N-3} \ket{0}_{N-2}
\left( \ket{0}_{N-1} \ket{+}_N + \ket{1}_{N-1} \ket{-}_N \right) \nonumber \\
+& \frac{1}{\sqrt{2}}\ket{\phi_1}_{N-3} \ket{1}_{N-2} 
\left( \ket{0}_{N-1} \ket{+}_N - \ket{1}_{N-1} \ket{-}_N \right)
\end{align}    
for $\ket{\pm} = \tfrac{1}{\sqrt{2}}(\ket{0} \pm \ket{1})$, where the
subscript labels the qubit, and the terms for the $(N-3)$ qubits are denoted
by $\ket{\phi_i}$.  The state $\ket{\Psi_{N+1}}$ can be generated by attaching
a qubit in $\ket{+}$ to $\ket{\Psi_{N}}$ by performing a CPHASE operation.  If
this operation fails, 
state~\eqref{eq:clus} becomes a mixed state.  However, $\ket{\Psi_{N - 2}}$
can be recovered from this state by measuring the $(N-1)$th qubit in the
computational basis and performing an operation on the
$(N-2)$nd qubit dependent on the measurement result.  The average number of
qubits attached to the cluster state by $m$ CPHASE operations is $(3P - 2)m$,
which grows if $P > 2/3$.   So, cluster states consisting of DF logical qubits
inside separate cavities can be prepared to high-fidelity using linear optics. 
\section{Conclusion}
We have proposed a general method to prepare a maximally
entangled DFS state in a linear array of dipole-coupled qubits that, although
relevant to any system of dipole-coupled qubits, in the
light of recent experimental progress, is directly applicable to a system of NV
centers in diamond.  The preparation is possible to high-fidelity, and can be
performed quickly relative to the linewidth of the entangled state.  We showed
how to rotate a logical qubit to high-fidelity within a DFS without leakage,
and quickly relative to the combined decay of the logical states.  Similar to
the preparation, we showed how to read out a logical state with high-fidelity.
We then described two methods to perform two-qubit operations, one of which
enables cluster-state preparation in a system of spatially separated DF
dipole-coupled qubits.  
\section*{Acknowledgments}
We especially thank Karl-Peter Marzlin for many helpful discussions. We
also thank Jim Cresser, Barry Sanders, and Jason Twamley for
comments on the manuscript.  This work was supported by Macquarie University.


\begin{thebibliography}{99}
\bibitem{Pal96Duan97Zan97aLidar98Knill00}
{G.M. Palma, K.-A. Suominen, and A. K. Ekert}, {Proc. Roy. Soc. London
  Ser. A}, \textbf{452}, 567 (1996);
{L.-M Duan and G.-C. Guo}, \prl \textbf{79},  1953  (1997);
{P. Zanardi and M. Rasetti}, {\it ibid}. \textbf{79},  3306  (1997);
{D.A. Lidar, I. L. Chuang, and K. B. Whaley}, {\it ibid}. \textbf{81},  2594 (1998);
{E. Knill, R. Laflamme, and L. Viola}, {\it ibid}. \textbf{84}, 2525 (2000). 
\bibitem{Be00Fe02Zan98}
{A. Beige, D. Braun, B. Tregenna, and P. L. Knight}, \prl \textbf{85}, 
1762 (2000);
{M. Feng and X. Wang}, \pra \textbf{65}, 044304 (2002);
{P. Zanardi and F. Rossi}, \prl \textbf{81}, 4752 (1998).
\bibitem{Pet02}
{D. Petrosyan and G. Kurizki}, \prl \textbf{89}, 207902 (2002).
\bibitem{Kem01}
{J. Kempe, D. Bacon, D. A. Lidar, and K. B. Whaley}, \pra \textbf{63}, 
042307 (2001).
\bibitem{Kiel01}
{D. Kielpinski \emph{et al}.}, Science \textbf{291},  1013  (2001).
\bibitem{Kw00}
{P. G. Kwiat, A. J. Berglund, J. B. Altepeter, and A. G. White}, Science
\textbf{290}, 498 (2000).
\bibitem{Bou04}
{M. Bourennane, M. Eibl, S. Gaertner,  C. Kurtsiefer, A. Cabello, and
  H. Weinfurter}, \prl \textbf{92}, 107901 (2004).
\bibitem{Moh03}
{M. Mohseni,  J. S. Lundeen, K. J. Resch, and A. M. Steinberg}, \prl
\textbf{91}, 187903 (2003). 
\bibitem{Lo01}
L. Viola {\it et al}., Science \textbf{293}, 2059 (2001).
\bibitem{Da94}
G. Davies, {\it Properties and growth of diamond} (IEE/INSPEC, London, 1994),
Vol. \textbf{9}.  
\bibitem{Je04}
F. Jelezko \emph{et al}., \prl \textbf{93}, 130501 (2004).
\bibitem{Ho05}
M. Howard \emph{et al}., New J. Phys. \textbf{8}, 33 (2006).
\bibitem{Me05Ra05}
J. Meijer \emph{et al}., {Appl. Phys. Lett.} \textbf{87}, 261909 (2005);
J.R. Rabeau \emph{et al}., {\it ibid}. (to be published).
\bibitem{Ni03Ni03a}
A.P. Nizovtsev \emph{et al}., {Opt. Spectrosc.} \textbf{94}, 848 (2003);
{Physica B} \textbf{340}, 106 (2003). 
\bibitem{Dr99}
A. Drabenstedt \emph{et al}., \prb \textbf{60}, 11503 (1999).
\bibitem{San06}
C. Santori \emph{et al}., Opt. Express \textbf{14}, 7986 (2006).
\bibitem{Mil74}
{P. W. Milonni and P. L. Knight}, \pra \textbf{10}, 1096 (1974).
\bibitem{Car93}
{H.J. Carmichael},  {\em
  {An Open Systems Approach to Quantum Optics}} (Springer-Verlag,
  Berlin, 1993). 
\bibitem{Dicke}
R.H. Dicke, Phys. Rev. \textbf{93}, 99 (1954).
\bibitem{Clem03}
{J. P. Clemens, L. Horvath, B. C. Sanders, and H. J. Carmichael}, \pra
\textbf{68}, 023809 (2003).
\bibitem{Bela69Lehm70iLehm70iiArg70}
A.A. Belavkin \emph{et al}., {Sov. Phys. JETP} \textbf{56}, 264 (1969);
G.S. Agarwal, \pra {\textbf 2}, 2038 (1970); 
R.H. Lehmberg, {\it ibid}. {\textbf 2}, 883 (1970); {\textbf 2} 889 (1970).
\bibitem{Lind}
{G. Lindblad}, Commun. Math. Phys. \textbf{48}, 119 (1976) 
\bibitem{Me05a}
J. Meijer  \emph{et al}., Appl. Phys. A \textbf{83}, 321 (2006).
\bibitem{Niel00}
{M.A. Nielsen and I.L. Chuang}, {\it Quantum Computation and Quantum
Information} (Cambridge University Press, Cambridge, England, 2000).
\bibitem{newfocus}
See, e.g., the specifications published by the manufacturers New Focus at
http://www.newfocus.com.
\bibitem{Bri01Rau01}
{H.J. Briegel and R. Raussendorf}, \prl \textbf{86}, 910 (2001);
{R. Raussendorf and H.J. Briegel}, {\it ibid}. \textbf{86}, 5188 (2001).
\bibitem{Ch05}
J. Cho and H.-W. Lee, \prl \textbf{95}, 160501 (2005).
\end{thebibliography}
\end{document}